\documentclass[manuscript,aps]{revtex4}
 \usepackage{graphicx}

 \def\cm{cm$^{-1}$}
 \def\Tc{$T_{\rm c}$}
 \def\CuO2{CuO$_2$}
 \def\Sm214{Sm$_2$CuO$_4$}
 \def\Sr112{SrCuO$_2$}
 \def\SrLnx{Sr$_{\rm 1-x}Ln_{\rm x}$CuO$_2$}
 \def\SrLn{Sr$_{0.9}Ln_{0.1}$CuO$_2$}
 \def\SrLa{Sr$_{0.9}$La$_{0.1}$CuO$_2$}
 \def\SrGd{Sr$_{0.9}$Gd$_{0.1}$CuO$_2$}
 \def\SrSm{Sr$_{0.9}$Sm$_{0.1}$CuO$_2$}

 \def\CaSr{Ca$_{0.86}$Sr$_{0.14}$CuO$_2$}
 
 \def\Tl-n{Tl$_2$Ba$_2$Ca$_{n-1}$Cu$_n$O$_{4+2n}$}

\begin{document}

 \draft
 \title{Optical phonons of superconducting infinite-layer 
	compounds \SrLn$~$($Ln$ = La, Gd, Sm)}
 \author{Mi-Ock Mun, Young Sup Roh, Kibum Kim, and Jae Hoon Kim}
 \affiliation{Institute of Physics and Applied Physics and Department of Physics \\
 		Yonsei University, Seoul 120-749 Korea}
 \author{C. U. Jung, J. Y. Kim, Min-Seok Park, Heon-Jung Kim, and Sung-Ik Lee}
 \affiliation{National Creative Research Initiative Center for Superconductivity\\
		Department of Physics \\
		Pohang University of Science and Technology \\
		Pohang 790-784 Korea}

\begin{abstract}

We have identified the optical phonon modes
of the superconducting infinite-layer compounds
\SrLn$~$($Ln$ = La, Gd, Sm; $T_c$ = 43 K for all) 
from their infrared reflectivity spectra 
obtained with a Fourier-transform infrared spectrometer
on high-quality high-purity samples synthesized 
under high pressure.
The La- and the Gd-doped compounds exhibited
only four (2$A_{\rm 2u}$+2$E_{\rm u}$)
out of the five (2$A_{\rm 2u}$+3$E_{\rm u}$) infrared-active phonon modes
predicted by a group theoretical analysis
whereas the Sm-doped compound exhibited all five modes.
We propose the atomic displacement pattern for each phonon mode
based on reported lattice dynamics calculations
and through comparison with the phonon modes 
of other single-layer high-$T_{\rm c}$ cuprate superconductors.

\vspace{3mm}
\small{PACS ; 74,74.25.-q,74.25.Gz,74.25.Kc}
\pacs{74,74.25.-q,74.25.Gz,74.25.Kc}
\vspace{5mm}
\end{abstract}
\maketitle

\newpage
\section{Introduction}

 The superconducting infinite-layer cuprate \SrLnx$~$($Ln$ = Lanthanide element),
 which consists solely of CuO$_2$ planes separated by metal ions,
 embodies the essential and minimal structure
 underlying all high-$T_{\rm c}$ cuprate superconductors.
 It provides us with a good opportunity
 to understand the mechanism of high-$T_{\rm c}$ superconductivity
 by isolating the intrinsic properties of the CuO$_2$ planes
 without complications arising from complex crystal structures.
 For this reason, since its discovery in 1988 \cite{Siegrist,Smith},
 much effort was made to measure and understand
 the basic physical properties of this compound 
 \cite{Wang,Norton,Ikeda,Jones,Feenstra1,Matsumoto}.
 However, the difficulties in obtaining high-quality 
 infinite-layer compounds so far,
 mostly due to impurities and/or multiphases,
 have made it hard to obtain reliable experimental data
 and to test currently available theoretical models.
 This is also the case with the infrared spectroscopic measurement,
 which is one of the basic probes of the electrons and phonons in solids.
 Previous reports by Burns {\em et al.} \cite{Burns}
 and Tajima {\em et al.} \cite{Tajima} on \CaSr,
 a nonsuperconducting isostructural compound,
 and those by Zhou {\em et al.} \cite{Zhou}
 and Er {\em et al.} \cite{Er} on superconducting \SrLn$~$contradict 
 one another in their mode assignments
 for the infrared-active phonons
 while the lattice dynamics calculations by Agrawal \cite{Agrawal}
 and Koval \cite{Koval}
 do not completely agree with experiment.

 In this paper, we report on our infrared reflectivity measurements
 on high-quality high-purity samples of the superconducting infinite-layer compounds
 \SrLn$~$($Ln$ = La, Gd, Sm; $T_c$ = 43 K for all) 
 synthesized under high pressure
 and propose our mode assignments for the optical phonons. 
 
\section{Experimental}

 Our samples of the superconducting infinite-layer compounds
 \SrLn$~$($Ln$ = La, Gd, Sm) were 
 synthesized by a high-pressure synthesis method \cite{silee1,silee2,silee3}.
 The zero-field-cooled measurement of the magnetic susceptibility
 clearly showed a sharp transition at \Tc$~$and
 indicated the Meissner fraction greater than 80 \%.
 The \Tc's are 43 K for all the samples studied,
 where the transition temperature 
 was determined by the criterion of 10 \% onset of superconductivity
 in the low-field magnetization measurements.

 Normal-incidence reflectivity measurements were carried out 
 with a Fourier-transform spectrometer (Bruker IFS 113v)
 in the spectral range of 40 - 5000 \cm$~$and
 with a grating spectrometer (Varian Cary 5G)
 from 4000 to 50000 \cm.
 
\section{Results and Discussion}

 The reflectivity spectra of the infinite-layer compounds
 \SrLn$~$($Ln$ = La, Gd, Sm) are shown in Fig. \ref{IL-R}.
 At low frequencies, the reflectivity curves show
 typical metallic response of the free carriers
 and the infrared-active phonon modes appear as peaks
 in the 50 - 700 \cm$~$range.
 We calculated the optical conductivity
 via a Kramers-Kronig analysis
 to identify the transverse-optical phonon modes
 (Fig. \ref{IL-S}).
 Our main objectives here are to determine 
 the center frequencies of the optical phonon modes
 and to propose the atomic displacement patterns 
 consistent with the available experimental and computational data.

 A group theoretical analysis predicts 
 five infrared-active optical phonons,
 two $A_{\rm 2u}$ modes and three $E_{\rm u}$ modes,
 for the space group (P4/mmm ; $D_{\rm 4h}^1$)
 relevant to the tetragonal structure of the infinite-layer compound.
 In \SrLa,
 only four modes were detected 
 at 152, 275, 354, and 559 \cm.
 The highest two modes at 354 and 559 \cm$~$were
 assigned to $E_{\rm u}$ modes
 in view of previous polarized measurements 
 ($\vec{\rm E} \perp {\bf c}$)
 on \CaSr$~$\cite{Tajima},
 which is a nonsuperconducting isostructural compound.
 The remaining $E_{\rm u}$ phonon is
 expected at about 230 \cm$~$but 
 is not detected most probably due to screening by free carriers.
 The two lowest modes at 152 and 275 \cm$~$of
 \SrLa$~$must then be $A_{\rm 2u}$ modes.
 It is quite natural for the $A_{\rm 2u}$
 modes to dominate the reflectivity spectra of polycrystalline samples
 \cite{Huml}.
 Our mode assignments agree 
 with the lattice dynamics calculation on \CaSr$~$by
 Koval {\em et al.} \cite{Koval}
 at least in the symmetry of the modes 
 although the actual frequencies and displacement patterns 
 are not completely consistent.
 \SrGd$~$shows similar phononic behavior
 with its second $E_{\rm u}$ mode more pronounced.
 As can be seen in Fig. \ref{IL-R} and Fig. \ref{IL-S},
 the \SrSm$~$sample is of relatively poor quality,
 showing in lower and more smeared reflectivity
 compared to \SrLa$~$and \SrGd.
 Nevertheless, all the five phonon modes were observed
 at 161, 208, 280, 355, and 561 \cm, respectively.
 The lowest $E_{\rm u}$ mode, not detected in the La- and the Gd-doped compounds,
 now appears at 208 \cm, which is consistent with the aforementioned
 polarized data on \CaSr$~$ ($\vec{\rm E} \perp {\bf c}$) \cite{Tajima}.

 Our assignment of the atomic vibration patterns for the infinite-layer compound
 (Fig. \ref{IL-pattern})
 is based on the analogy with the case of $n$-type superconductors
 of the $T^{\prime}$ structure \cite{Heyen},
 which also contain single CuO$_2$ planes without any apical oxygen.
 For \SrLa,
 the two $E_{\rm u}$ modes
 at 354 \cm$~$and 559 \cm$~$are assigned to the 
 Cu-O bending and the Cu-O stretching motions in the $ab$-plane,
 respectively.
 These two planar vibrations are commonly observed 
 in other cuprates at similar eigenfrequencies
 because they are associated with the CuO$_2$ plane itself
 and hence are not largely affected by the nature of
 the insulating or charge-reservoir blocks.
 The remaining $E_{\rm u}$ mode, which is detected only in the Sm-doped compound,
 is most likely to arise from the out-of-phase motion of
 Sr/$Ln$ metal ion sheets against \CuO2$~$planes in the planar direction
 (the external mode).
 The lowest $A_{\rm 2u}$ mode is assigned to
 the out-of-phase vibration of metallic Sr/$Ln$ ions against Cu ions along the $c$ axis.
 This mode shifts more or less with $Ln$ replacement.
 The second $A_{\rm 2u}$ mode at 275 \cm$~$for
 \SrLa$~$is assigned to the dimpling motion of oxygen 
 and copper ions along the $c$ axis
 without involving the motion of heavy metallic ions,
 which is consistent with the robustness of this mode against ion substitution.
 In Table \ref{tbl-1}, we summarized our assignment of the phonon modes.

\section{Conclusion}
 The five optical phonons, 2$A_{2u}$ + 3$E_u$, 
 of the high-purity high-quality
 superconducting infinite-layer compounds \SrLn$~$were
 identified and assigned on the basis of
 a group theoretical analysis, reported lattice dynamics calculations
 and comparison with other single-layer cuprate materials.
 
\acknowledgements

MOM was supported by Basic Research Program for the Women Scientists
of the Korea Science \& Engineering Foundation
for this work.
JHK was supported by BK21 Project of Ministry of Education.
SIL was supported by the Creative Initiative Research Program of 
the Ministry of Science and Technology of Korea and KOSEF
(Contract Number 95-0702-03-03-3).

\newpage
\begin{table}[h]
\caption{The phonon frequencies and the proposed atomic vibration pattern
	of the superconducting infinite-layer compound \SrLn$~$.}
\label{tbl-1}
\begin{center}
 \begin{tabular}{cc|ccccccc|cl}
   \hline\hline
   &&\multicolumn{6}{c}{~~Frequency (\cm)}&& \\
   \multicolumn{1}{c}{Mode} &
   \multicolumn{1}{c}{\hspace{7mm}}&
   \multicolumn{1}{|c}{\hspace{7mm}}&
   \multicolumn{1}{c}{$~~$La} &
   \multicolumn{1}{c}{\hspace{7mm}} &
   \multicolumn{1}{c}{$~~~$Gd} &
   \multicolumn{1}{c}{\hspace{7mm}}&
   \multicolumn{1}{c}{$~~~$Sm} &
   \multicolumn{1}{c}{\hspace{10mm}}&
   \multicolumn{1}{|c}{\hspace{10mm}}&
   \multicolumn{1}{c}{Atomic Vibration Pattern} \\
   \hline
   $~~A_{2u}$(1)& & & $~~$152 & & $~~~$148 & & $~~~$161 & & & 
   $~$out-of-phase vibration of heavy ion against Cu along $c$ axis \\
   $~~E_{u}$(1) & & & $~~$ -	& & $~~~$-   & & $~~~$208 & & & 
   $~$in-plane external motion (\CuO2$~$plane against Sr/$Ln$ sheet)$~$ \\
   $~~A_{2u}$(2)& & & $~~$275 & & $~~~$275 & & $~~~$280 & & & 
   $~$Cu-O dimpling motion along $c$ axis  \\
   $~~E_{u}$(2) & & & $~~$354 & & $~~~$353 & & $~~~$355 & & & 
   $~$in-plane Cu-O bending mode\\
   $~~E_{u}$(3) & & & $~~$559 & & $~~~$562 & & $~~~$561 & & & 
   $~$in-plane Cu-O stretching mode\\
   \hline\hline
 \end{tabular}
\end{center}
\end{table}

\newpage
\begin{figure}[h]
 \caption{The reflectivity of \SrLn$~$measured at normal-incidence :
	  solid line for \SrLa, dash-dot line for \SrGd, and dashed line for \SrSm.} 
 \label{IL-R}
\end{figure}

\begin{figure}[h]
 \caption{The real part of the optical conductivity of \SrLn$~$obtained by a
	  Kramers-Kronig analysis :
	  solid line for \SrLa, dash-dot line for \SrGd, and dashed line for \SrSm.
	 } 
 \label{IL-S}
\end{figure}

\begin{figure}[h]
 \caption{The atomic vibration patterns of \SrLn$~$with
 the tetragonal structure of P4/mmm(D$^1_{4h}$) :
 black circle for Cu, white circle for O, and gray circle for Sr/$Ln$.} 
 \label{IL-pattern}
\end{figure}

\end{document}